\documentclass[aps,prd,twocolumn,groupedaddress,eqsecnum]{revtex4}

\usepackage{bm,amsmath}

\begin{document}

\title{Thermal field theory and generalized light front quantization}


\author{H.  Arthur Weldon}

\affiliation{Department of Physics, West Virginia University, Morgantown,
West Virginia 26506-6315}


\begin{abstract}
The dependence of thermal field theory on the surface of quantization
and on the velocity of the heat bath is investigated by working in 
general coordinates  that are arbitrary linear
combinations of the Minkowski coordinates. In the general coordinates the metric
tensor $g_{\overline{\mu\nu}}$ is non-diagonal. The Kubo, Martin,
Schwinger condition requires periodicity in thermal correlation
functions when the temporal variable
 changes by an amount
$-i\big/(\,T\sqrt{g_{\overline{00}}}\,)$.
Light front quantization fails since $g_{\overline{00}}=0$, however
various related quantizations are possible.

\end{abstract}


\pacs{11.10.Wx, 12.38.Mh}



\maketitle

\section{introduction}

\subsection{The light front and thermal field theory}

For many years light front quantization has been applied to deep inelastic scattering
and the Wilson operator product expansion. 
More recently it has been used in QCD to compute hadron structure
\cite{b,c,d,e}. 
Light front quantization brings both 
conceptual and computational simplifications  to certain  hadronic
processes \cite{PD,rev}. 

Much of the computational simplification occurs because  the mass shell condition
expressed in terms of the  
 momentum variables $p^{\pm}=(p^{0}\pm p^{3})/\!\sqrt{2}$ is
\begin{subequations}
\begin{equation}
p^{-}={\bm{p}_{\bot}^{2}+m^{2}\over 2p^{+}}.\label{LCdisrel}
\end{equation}
The operator $P^{-}=(P^{0}-P^{3})/\sqrt{2}$  plays
the role of the Hamiltonian in that it generates the evolution of the fields in
the coordinate
$x^{+}=(x^{0}+x^{3})/\sqrt{2}$
\begin{equation}
\big[P^{-},\phi(x)\big]=-i{\partial\phi\over\partial x^{+}}.
\end{equation}

Recently Brodsky suggested \cite{SB2} that the computational advantages
of quantizing in light front coordinates
might carry over to statistical mechanics and thermal field theory
done on the light front and proposed as the appropriate partition
function 
\begin{equation}
{\rm Tr}\big[\exp\big\{- P^{-}/T_{\rm LC}\big\}\big].\label{ZLC}
\end{equation}\end{subequations}
The relation of $T_{LC}$ to the usual invariant temperature
$T$ was unspecified. 

The suggestion was pursued by
Alves, Das, and Perez \cite{ADP}. For a free field
theory they found an immediate problem that comes from the vanishing of the on-shell energy,
Eq. (\ref{LCdisrel}),  as $p_{+}\to\infty$.
The breakdown is not specific to
the canonical ensemble. In the microcanonical ensemble the entropy 
is a measure of the multiparticle  phase space available to a gas of particles whose
total energy is fixed.  To any configuration with a fixed value of the total $P^{-}$
one can add any number of zero-energy particles each having an infinite
value of $p^{+}$. The entropy of such a configuration diverges.

Alves, Das, and Perez \cite{ADP} showed that a thermal average performed in the rest
frame but using light-front variables does work.  Using
$P^{0}=(P^{+}+P^{-})/\sqrt{2}$ the partition function is
\begin{equation}
{\rm
Tr}\Big[\exp\Big\{-(P^{+}+P^{-})/\sqrt{2}\,T\Big\}\Big],\label{ZADP}
\end{equation}
instead of Eq. (\ref{ZLC}).
They performed one-loop calculations of the  self-energy
in  scalar field theories with either a $\phi^{4}$ interaction or a $\phi^{3}$
interaction.  The final
results for both calculations were exactly the same as the conventional answers.

\subsection{Thermal field theory in generalized coordinates}

In order to explore the various possible options it is most efficient
to consider quantization in a general set of space-time coordinates
and later examine light front quantization as a special case. 
The metric signature is $(+,-,-,-)$. 

The conventional approach is to impose quantization conditions on fields  at a
fixed value of
$x^{0}$. The operator that generates time evolution  is 
$P_{0}$. The partition function is
\begin{equation}
{\rm Tr}\big[\exp\big\{- P_{0}/T\big\}\big],
\end{equation}
as is appropriate for a heat bath at
rest. If this  system is viewed from a Lorentz frame with velocity $v$ and
four-velocity $u^{\alpha}=(\gamma, 0,0,\gamma v)$, then the quantization will be
at a fixed value of $x^{\alpha}u_{\alpha}$; the evolution operator will be
$P_{\alpha}u^{\alpha}$; and the partition function will be
${\rm Tr}[\exp(-\beta P_{\alpha}u^{\alpha})]$. The fact that $u^{\alpha}$ serves as
both the vector normal to the surface of quantization and the velocity vector of the
heat bath is a unique feature of Lorentz boosted coordinates and is not be true
in more general coordinates.

The present paper  investigates the dependence of thermal field theory
on the surface of quantization and on the velocity of the heat bath.
Subsequent analysis will show that
for any  vector $n_{\alpha}$  that is time-like or light-like, 
it is possible to quantize at a fixed value of $n_{\alpha}x^{\alpha}$,
\begin{subequations}\label{nu}\begin{equation}
\big[\pi(x), \phi(0)\big]\,\delta(n_{\alpha}x^{\alpha}-c)
=-i\,\delta^{4}(x), 
\end{equation}
and employ the partition function
\begin{equation}
{\rm Tr}\big[\exp\big\{- P_{\alpha}u^{\alpha}/T\big\}\big],
\end{equation}
appropriate to a  heat bath
moving with four-velocity $u^{\alpha}=(\gamma,0,0,\gamma v)$,
provided  only that $n_{\alpha}$ and $u^{\alpha}$ satisfy
\begin{equation}
u^{\alpha}n_{\alpha}> 0.\label{recip}
\end{equation}\end{subequations}
The conventional rest frame choice is $n^{0}=u^{0}=1$ and $n^{j}=u^{j}=0$. A Lorentz boost
from the rest frame still gives
$n^{\alpha}=u^{\alpha}$. The equality of these two vectors is not general as various examples
will show. One can see that  light front quantization using the partition function in Eq.
(\ref{ZLC}) fails, even apart from the divergence issues mentioned previously,  because it
requires $v\to 1$ in order that
$P_{\alpha}u^{\alpha}$ be proportional to
$P_{-}$. This makes $T_{\rm LC}=T\sqrt{1-v^{2}}/\sqrt{2}\to 0$.  
In contrast, the successful approach of Alves, Das, and Perez \cite{ADP} in Eq.
(\ref{ZADP}) fits into the general framework above with
$n^{\alpha}=(1,0,0,1)$ and $u^{\alpha}=(1,0,0,0)$. 

The purpose of this paper is to  investigate thermal field
theory formulated in general coordinates that are arbitrary linear combinations of
the Cartesian $t, x, y, z$ and see if there are any computational advantages to
other formulations. The method is conservative in that the new coordinates are
restricted to be linear combinations of the Cartesian coordinates \cite{Fub},
a restriction which guarantees that physical consequences will be the same  for the following reason.
The Lagrangians of fundamental field theories are invariant under
arbitrary nonlinear coordinate transformations in accord with the principle
of equivalence and thus are trivially invariant under the  linear
coordinate transformations considered here. Physically interesting quantities
in thermal field theories are  either scalars, spinors, or tensors.
Whether calculated in Cartesian coordinates or more general linear coordinates,
the  scalar quantities should be identical and the spinor and tensor quantities
should be simply related by the chosen transformations.

Section II develops the formalism of thermal field theory in the general coordinates in order
to determine  the possible  choices for the surface of quantization and the velocity of the
heat bath. Section III treats several examples that are specifically related to light front
quantization and it may be read independently of Sec. II.  Section IV provides some conclusions.

\section{Thermal Field Theory in Oblique Coordinates}

\subsection{Transformed coordinates and metric}

From the Cartesian coordinates $x^{0}=t$, $x^{1}$, $x^{2}$, $x^{3}$ a new set of
coordinates $x^{\overline{0}}$, $x^{\overline{1}}$, $x^{\overline{2}}$,
$x^{\overline{3}}$ can be formed by taking  linear combinations: 
\begin{equation}
x^{\overline\mu}=A^{\overline\mu}_{\alpha}\,x^{\alpha},\label{A1}\end{equation}
where the $A^{\overline\mu}_{\alpha}$ are a set of 16 real constants. 
The notation used is due to Schouten \cite{Sc}. It expresses the fact that the
space-time point specified by the four-vector
$x$ does not change under a coordinate transformation; only the labels used to
indentify the components change.  For example, the light front choice is expressed
$x^{\overline{0}} =(x^{0}+x^{3})/\sqrt{2}$. 
It is also convenient that for $A^{\overline\mu}_{\alpha}$ the
superscript
$\overline\mu$ is neither a first index nor a second index since $\overline{\mu}$ 
cannot be confused with $\alpha$. 

It is misleading to refer to the general coordinate transformations as a change in the
reference frame. A change in the physical reference frame can only be done by rotations and Lorentz
boosts. 

The  16 real numbers
$A^{\overline{\mu}}_{\alpha}$ must have a nonzero determinant. 
To guarantee that $x^{\overline{0}}$ be a time coordinate, it is
necessary to require that the covariant Lorentz vector
$A^{\overline{0}}_{\alpha}$ be time-like or light-like:
\begin{equation}
A^{\overline{0}}_{\alpha}A^{\overline{0}}_{\beta}\,g^{\alpha\beta}\ge
0,\label{time1}
\end{equation}
where $g_{\alpha\beta}={\rm diag}(1,-1,-1,-1)$ is the Minkowski metric. 
It is sometimes useful to express the transformation matrix  in
terms of partial derivatives:
\begin{equation}
A^{\overline\mu}_{\alpha}={\partial x^{\overline\mu}\over\partial
x^{\alpha}}.
\end{equation}
The relation (\ref{A1}) can be inverted to find the Cartesian coordinates
$x^{\beta}$ as linear combinations of the new coordinates $x^{\overline\nu}$.
These partial derivatives are denoted
\begin{equation} 
A^{\beta}_{\overline\nu}={\partial x^{\beta}\over\partial
x^{\overline\nu}}.\label{Adef}
\end{equation}
The transformations satisfy
\begin{equation}
A^{\overline{\mu}}_{\alpha}\,A^{\alpha}_{\overline\nu}=\delta^{\overline\mu}
_{\overline\nu}
\hskip1cm
A^{\overline\mu}_{\alpha}\,A^{\beta}_{\overline\mu}
=\delta^{\beta}_{\alpha}.
\end{equation}

The differential length element in Cartesian coordinates is
$g_{\alpha\beta}\,dx^{\alpha}dx^{\beta}=(dt)^{2}\!-\!(dx)^{2}\!-\!(dy)^{2}\!-\!(dz)^{2}$.
In the new coordinates there is a new metric $g_{\overline{\mu\nu}}$ given by
\begin{equation}
g_{\overline{\mu}\overline{\nu}}=A^{\alpha}_{\overline\mu}\,
A^{\beta}_{\overline{\nu}}\,g_{\alpha\beta},
\end{equation}
which satisfies 
$g_{\alpha\beta}\,dx^{\alpha}dx^{\beta}=
g_{\overline{\mu\nu}}\,dx^{\overline{\mu}}dx^{\overline\nu}$. 
 If all 12 of the off-diagonal entries in
$g_{\overline{\mu\nu}}$ vanish, the new coordinates are orthogonal. This occurs,
for example, if  the $A$ are Lorentz transformations. In general the
$g_{\overline{\mu\nu}}$ are not diagonal and in general the new coordinates 
are oblique. It will be convenient
for later purposes to denote the determinant of the covariant metric by
\begin{equation}
g={\rm Det}\Big[ g_{\overline{\mu\nu}}\Big]<0.
\end{equation}
The Jacobian of the coordinate transformation
\[
dx^{0}dx^{1}dx^{2}dx^{3}=Jdx^{\overline{0}}dx^{\overline{1}}
dx^{\overline{2}}dx^{\overline{3}},\]
is therefore
\begin{equation}
J={\rm Det}\Big[A^{\alpha}_{\overline\mu}\Big]=\sqrt{-g}.
\end{equation}

It will also be necessary to use the contravariant metric $g^{\overline{\mu\nu}}$,
related to the Minkowski metric by
\begin{equation}
g^{\overline{\mu}\overline{\nu}}=A^{\overline\mu}_{\alpha}\,
A^{\overline\nu}_{\beta}\,g^{\alpha\beta}.
\end{equation}
The requirement in Eq. (\ref{time1}) can be stated as
\begin{equation}
g^{\overline{00}}\ge 0.\label{g001}
\end{equation}

\subsection{Four-momentum operators}

In the conventional quantization at fixed $x^{0}$ the momentum
operators $P_{\lambda}$ are the generators of space-time translations.
For a scalar field operator $\phi(x)$ they satisfy
\begin{displaymath}
\big[P_{\lambda},\phi\big]=-i{\partial\phi\over\partial x^{\lambda}}.
\end{displaymath}
The linear combination of these generators defined by
\begin{equation}
P_{\overline{\mu}}=A^{\lambda}_{\overline\mu}\,P_{\lambda},\label{Ptran}
\end{equation}
with $A^{\lambda}_{\overline\mu}$ given in Eq. (\ref{Adef}) will  satisfy
\begin{equation}
\big[P_{\overline{\mu}},\phi\big]=-i{\partial x^{\lambda}\over
\partial x^{\overline{\mu}}}\,{\partial\phi\over\partial x^{\lambda}}
=-i{\partial\phi\over\partial x^{\overline\mu}}.
\end{equation}
In particular, $P_{\overline{0}}$ generates the evolution in the variable
$x^{\overline{0}}$. Appendix B  shows that if one quantizes at fixed
$x^{\overline{0}}$ then the Hamiltonian is this same operator $P_{\overline{0}}$. 

Note that the temporal evolution is in the variable
$x^{\overline 0}=A^{\overline 0}_{\alpha}\,x^{\alpha}$ but the
evolution operator is in a different linear combination:
$P_{\overline 0}=A^{\alpha}_{\overline{0}}\, P_{\alpha}$.
These two vectors are reciprocal in the sense that
$A^{\overline 0}_{\alpha}\,A^{\alpha}_{\overline 0}=1$. 
Lorentz transformations  are special in that the metrics are the same,
$g_{\overline{\alpha\beta}}=g_{\alpha\beta}$ and the two vectors are the same,
$A^{\overline{0}}_{\alpha}=g_{\alpha\beta}\,A^{\beta}_{\overline{0}}$.

\subsection{Density operator}

Field theory at finite temperature requires a density operator and it is not
obvious exactly what the density operator should be when using general oblique
coordinates. For conventional quantization at fixed $x^{0}$ the density operator
has the form
\begin{equation}
\hat{\varrho}=\exp\big\{\!-\beta P_{\alpha}u^{\alpha}\big\},\label{Z0}
\end{equation}
where $u^{\alpha}$ is some four-velocity. Rewriting  this in general 
coordinates gives 
$\hat{\varrho}=\exp\Big\{-\beta
P_{\overline\sigma}\,u^{\overline\sigma}\,\Big\}.$
The appropriate value of $u^{\overline\sigma}$ is  undetermined but since the
Minkowski four-velocity satisfies $u^{\alpha}u^{\beta}g_{\alpha\beta}=1$, the velocity in
oblique coordinates must satisfy
\begin{equation}
u^{\overline\sigma}u^{\overline\mu}g_{\overline{\sigma\mu}}=1.
\end{equation}

\subsubsection{KMS relation}

With a partition function of the above form, the thermal Wightman
function is
\begin{displaymath}
{\cal D}_{>}(x)={\rm Tr}\big[\hat{\varrho}\,
\phi(x)\phi(0)\big]\big/{\rm Tr}[\hat{\varrho}\,].
\end{displaymath}
Consider a shift  the oblique time, i.e. $x^{\overline{0}}\to
x^{\overline{0}} +\delta x^{\overline{0}}$ with fixed values of $x^{\overline{1}},
x^{\overline{2}}, x^{\overline{3}}$. In terms of the Cartesian
coordinates 
$x^{\alpha}\to x^{\alpha}+\delta
x^{\alpha}$ with $\delta x^{\alpha}=A^{\alpha}_{\overline{0}}\,\delta x^{\overline{0}}$. Using
$P_{\alpha}\,\delta x^{\alpha}=P_{\overline{0}}\,\delta x^{\overline{0}}$ the shifted field operator
is
\begin{eqnarray}
\phi(x+\delta x)
=&&\exp(iP_{\alpha}\,\delta x^{\alpha})\phi(x)
\exp(-iP_{\alpha}\,\delta x^{\alpha})\nonumber\\
=&&\exp(iP_{\overline{0}}\;\delta x^{\overline{0}})\phi(x)\exp(-iP_{\overline{0}}\;\delta
x^{\overline{0}}).\nonumber
\end{eqnarray}
For evolution in imaginary values of $\delta x^{\overline{0}}$ to be equivalent to
thermal averaging requires that the density operator involve only $P_{\overline{0}}$ and
not $P_{\overline{j}}$. Therefore the spatial components of the contravariant velocity must vanish:
$u^{\overline{1}}=u^{\overline{2}}=u^{\overline{3}}=0$.  This describes a heat bath that is
at rest in the oblique coordinates. 
The normalized velocity vector is 
\begin{equation}
u^{\overline\sigma}=\Big({1\over\sqrt{g_{\overline{0}\overline{0}}}}
,0,0,0\Big).\label{vel}
\end{equation}
Note that this imposes a new requirement on the metric,
\begin{equation}
g_{\overline{00}}>0,\label{g002}\end{equation}
that is different than Eq. (\ref{g001}). 
This condition prevents standard light front quantization since $g_{\overline{00}}$ would vanish.
In all subsequent discussion it will be assumed that Eq. (\ref{g002}) is satisfied.
The density operator is
\begin{equation}
\hat{\varrho}=\exp\big\{-\beta P_{\overline{0}}\,\big/\!\sqrt{g_{\overline{00}}}\,\big\},
\label{Z1}
\end{equation}
or equivalently Eq. (\ref{Z0}) 
if the velocity vector is expressed in Cartesian coordinates:
\begin{equation}
u^{\alpha}=A^{\alpha}_{\overline\mu}\,u^{\overline{\mu}}=A^{\alpha}_{\overline{0}}
\,\big/\!\sqrt{g_{\overline{00}}}\ .\label{vel2}
\end{equation}
The vector normal to the quantization surface is $n_{\alpha}=A^{\overline{0}}_{\alpha}$ and
therefore $n_{\alpha}u^{\alpha}=1/\sqrt{g_{\overline{00}}}>0$ as stated in 
Eq. (\ref{recip}).

The density operator is used to define the thermal averaged Wightman function:  
\begin{equation}
{\cal D}_{>}(x)={\rm Tr}\big[\exp\big\{-\beta
P_{\overline{0}}\,/\!\sqrt{g_{\overline{00}}}\,\big\}
\phi(x)\phi(0)\big]\big/Z,
\end{equation}
where $Z={\rm Tr}[\hat{\varrho}]$ is the partition function.
Under imaginary shifts in the oblique time of the form $\delta
x^{\overline{0}}=i\alpha/\!\sqrt{g_{\overline{00}}}$, the behavior is
\begin{eqnarray}
&& {\cal D}_{>}\big(x^{\overline{0}}+i\alpha
\big/\!\sqrt{g_{\overline{00}}}\,,x^{\overline{j}}\,\big)\label{D>1}\\ 
&& ={\rm Tr}\bigg[\exp\bigg\{\!-\!
{(\beta+\alpha)P_{\overline{0}}\over \sqrt{g_{\overline{00}}}}\,\bigg\}
\phi(x)
\exp\bigg\{ {\alpha
P_{\overline{0}}\over\sqrt{g_{\overline{00}}}}\,\bigg\}\phi(0)\bigg]\big/Z.
\nonumber
\end{eqnarray}
The spectrum of $P_{\overline{0}}$ is bounded from below but will always have
arbitrarily large positive eigenvalues. If the two exponents are negative then
the infinitely large energies will be suppressed. In other words, (\ref{D>1}) is
analytic for $-\beta\le\alpha\le 0$. 
For the choice  $\alpha=-\beta $
or equivalently $\delta x^{\alpha}=-i\beta u^{\alpha}$ the result is
\begin{equation}
{\cal D}_{>}\big(x^{\overline{0}}-i\beta
\big/\sqrt{g_{\overline{00}}}\,,x^{\overline{j}}\,\big) ={\cal
D}_{>}\big(-x^{\overline{0}},-x^{\overline{j}}\,\big).
\end{equation}
This is the Kubo-Martin-Schwinger relation \cite{KMS} expressed in oblique coordinates.

\subsubsection{Tolman's law}

The dependence of the partition function (\ref{Z1}) on the combination $P_{\overline{0}}\big/
T\!\sqrt{g_{\overline{00}}}$ merits further discussion. One explanation comes from perturbation
theory in which at each order the eigenvalues of the operator $P_{\overline{0}}$ are the sum 
of single particle energies $p_{\overline{0}}$, each satisfying a mass shell condition
$g^{\overline{\mu\nu}}p_{\overline\mu}p_{\overline{\nu}}=m^{2}$, for various masses. Among the
allowed coordinate transformations are scale transformations. Rescaling the contravariant time by a
factor $\lambda$, as in
$x^{\overline{0}}\to \lambda\, x^{\overline{0}}$, would rescale the covariant energy by
$p_{\overline{0}}\to p_{\overline{0}}/\lambda$. The combination
$P_{\overline{0}}/T$ would not be invariant under this transformation. However, the scale
transformation changes the covariant metric, $g_{\overline{00}}\to g_{\overline{00}}/\lambda^{2}$,
and the combination $P_{\overline{0}}\big/
T\!\sqrt{g_{\overline{00}}}$ is invariant.

There is another way to understand why the partition function (\ref{Z1}) depends on the combination
$T\!\sqrt{g_{\overline{00}}}$. A condition necessary for thermal equilibrium in inertial
coordinates is that the temperature should be uniform in space and time. 
Tolman \cite{RCT} investigated the conditions for thermal equilibrium in 
a gravitational field and showed that a temperature gradient is necessary to prevent
the flow of heat from regions of higher gravitational potential to regions of lower gravitational
potential. The quantitative result is summarized by the statement that the product
$T\!\sqrt{g_{\overline{00}}}$ must be constant and is known at Tolman's law \cite{WI}. 

An
alternative derivation is given by Landau and Lifshitz \cite{LL}, who discuss how to compute the
entropy in the microcanonical ensemble in a general curvilinear coordinate system (with or without
gravity). In the microcanonical ensemble, the temperature is computed by differentiating the 
entropy with respect to the energy and this leads to the result that $T\!\sqrt{g_{\overline{00}}}$
must be constant.

\subsubsection{Covariant density operator}

For later purposes it is convenient to express the density operator (\ref{Z1}) in 
a covariant form in terms of the conserved energy-momentum operator 
$T^{\overline\nu}_{\;\cdot\;\overline{\mu}}$.   The generators of  translations are
\[ P_{\overline\mu}=\sqrt{-g}\int dx^{\overline{1}}dx^{\overline{2}}
dx^{\overline{3}}\,T^{\overline{0}}_{\;\cdot\;\overline{\mu}}.
\]
Rewrite this in terms of the energy-momentum operator in  Cartesian coordinates,
$T^{\alpha}_{\;\cdot\;\lambda}$ as
\begin{equation}
P_{\overline\mu}
=\int dS_{\alpha}\,
T^{\alpha}_{\;\cdot\;\lambda}\,A^{\lambda}_{\overline\mu},
\end{equation}
in accordance with Eq. (\ref{Ptran}). 
The three-dimensional, differential  surface element  is
\begin{equation}
dS_{\alpha}
=\sqrt{-g}\,A^{\overline{0}}_{\alpha}\, dx^{\overline{1}}
dx^{\overline{2}}dx^{\overline{3}}.\label{dS}
\end{equation}
This surface of quantization  is orthogonal to  three contravariant: vectors
$A^{\alpha}_{\overline{j}}\,dS_{\alpha}=0$ for
$\overline{j}=\overline{1},\overline{2},\overline{3}$. 
Contracting $P_{\overline\mu}$ with $u^{\overline\mu}$  and using Eq. (\ref{vel}) gives
the density operator in covariant form:
\begin{equation}
\hat{\varrho}
=\exp\Big\{-\beta\int dS_{\alpha} T^{\alpha}_{\;\cdot\;\lambda}
u^{\lambda}\Big\}.\label{Z2}
\end{equation}

\subsection{Thermal field theory in real time}

To quantize a field theory at a fixed value of $x^{\overline{0}}$ is straightforward but there are
some unfamiliar aspects that originate from the oblique metric
$g_{\overline{\mu\nu}}$ being non-diagonal. Appendix B performs the explicit
quantization for a scalar field theory. 
 It is most natural to  deal with contravariant
space-time coordinates
$x^{\overline\mu}$ and covariant momentum variables $p_{\overline\mu}$
so that the solutions of the field equations are superpositions of plane waves of
the form $\exp(\pm ip_{\overline\mu}x^{\overline\mu})$.  For spinor particles the Dirac matrices are $\gamma^{\overline\mu}=
A^{\overline\mu}_{\alpha}\gamma^{\alpha}$ and satisfy
$\{\gamma^{\mu},\gamma^{\nu}\}=2g^{\overline{\mu\nu}}$.
For gauge boson propagators there are  natural generalizations of Coulomb gauge,
axial gauge, and covariant gauges.
 This section will summarize some
familiar results but  expressed in oblique coordinates.  Either canonical quantization 
or functional integration may be used \cite{Das,LeBellac}.

\subsubsection{Propagators}

At zero
temperature the free propagator for a scalar field  is
\begin{equation} D(p_{\overline{0}}, p_{\overline{j}})={1\over g^{\overline{\mu\nu}}
p_{\overline\mu}p_{\overline\nu}-m^{2}+i\epsilon}\;.\label{prop1}
\end{equation}
The integration measure over loop momenta is
\begin{equation}
\int{dp_{\overline{0}}dp_{\overline{1}}dp_{\overline{2}}dp_{\overline{3}}
\over\sqrt{-g}(2\pi)^{4}}.
\end{equation}
At non-zero temperature the propagator has the usual $2\times 2$ matrix structure
\cite{Das,LeBellac}. The Bose-Einstein or Fermi-Dirac functions
become
\begin{equation}
n={1\over\exp(\beta |p_{\overline{0}}|/\sqrt{g_{\overline{00}}})\mp 1}.
\end{equation}

As discussed later, the most interesting possibility is to choose a coordinate transformation such
that
$g^{\overline{00}}=0$. This makes the denominator of  the propagator linear in
$p_{\overline{0}}$ and therefore there is only one pole in (\ref{prop1}).

If $g^{\overline{00}}\neq 0$
then the propagator has poles at two values of $p_{\overline{0}}$ but  the positive
and  negative values of $p_{\overline{0}}$ will  have different magnitudes. 
It  is  sometimes convenient to  express  the propagator in a mixed form, in terms of 
the covariant energy $p_{\overline{0}}$ but the contravariant momenta
$p^{\overline{j}}$.  To do this, use the identity  
\begin{eqnarray}
g_{\overline{\mu\nu}}p^{\overline\mu}p^{\overline\nu}=&&
{1\over g_{\overline{00}}}\big(g_{\overline{00}}\,p^{\overline{0}}
+g_{\overline{0i}}\,p^{\overline{i}}\,\big)^{2}
+\bigg(g_{\overline{ij}}-{g_{\overline{oi}}g_{\overline{0j}}\over g_{\overline{00}}}
\bigg)p^{\overline{i}}p^{\overline{j}}\nonumber\\
=&&  {(p_{\overline{0}})^{2}\over g_{\overline{00}}}-\gamma_{\overline{ij}}
p^{\overline{i}}p^{\overline{j}}\nonumber
\end{eqnarray}
where $\gamma_{\overline{ij}}$ is given by
\begin{equation}
\gamma_{\overline{ij}}=-g_{\overline{ij}}+{g_{\overline{oi}}g_{\overline{0j}}\over
g_{\overline{00}}}.\label{gamma}
\end{equation}
The determinant  is ${\rm Det}[\gamma_{\overline{ij}}]
=-g/g_{\overline{00}}>0$ \cite{com1}.
Define the effective energy
\begin{equation}
E=\Big[m^{2}+\gamma_{\overline{ij}}\,p^{\overline{i}}p^{\overline{j}}\Big]^{1/2}.
\end{equation}
The same propagator in these variables is
\begin{equation}
D(p_{\overline{0}}, p^{\overline{j}})={g_{\overline{00}}\over
(p_{\overline{0}})^{2}-g_{\overline{00}}\,E^{2}+i\epsilon}
\end{equation}
The poles are now at $p_{0}=\pm\sqrt{g_{\overline{00}}}\,E$. 
The Bose-Einstein or Fermi-Dirac functions
become
\begin{equation}
n= {1\over \exp(\beta E)\mp 1}.
\end{equation}
The integration  over covariant energy and  contravariant three-momenta is
\begin{equation}{\sqrt{-g}\over  g_{\overline{00}}}
\int{dp_{\overline{0}}\,dp^{\overline{1}}dp^{\overline{2}}dp^{\overline{3}}
\over (2\pi)^{4}}.
\end{equation}

\subsubsection{Statistical mechanics}

The partition function provides a direct calculation  
 of the thermodynamic
functions via the  Helmholtz free energy  $F$:
\begin{equation}
\exp\Big\{-\beta F/\sqrt{g_{\overline{00}}}\Big\}
={\rm Tr}\Big[\exp\big\{-\beta P_{\overline{0}}/\sqrt{g_{\overline{00}}}\big\}
\Big].\label{SM1}
\end{equation} 
The free energy $F(T,V)$ then allows computation of the pressure and entropy:
\begin{subequations}\begin{eqnarray}
P=&&-\bigg({\partial F\over\partial V}\bigg)_{T}\label{P}\\
S=&&-\bigg({\partial F\over \partial T}\bigg)_{V},\label{S}
\end{eqnarray}\label{thermo}\end{subequations}
and the energy is $U=F+TS$. 

\subsubsection{Covariant statistical mechanics}

The prescriptions in Eq. (\ref{thermo}) can be derived rather elegantly from the partition  function 
if it is  formulated covariantly \cite{WI,AW}. 
In global thermal equilibrium considered here, the temperature $T$ and the
velocity
$u^{\overline\mu}$ of the heat bath are independent of space-time and 
so there is no heat flux or viscosity. The thermal average of the energy-momentum
operator has the perfect fluid form:
\begin{equation}
{{\rm Tr}\Big[\hat{\varrho}\,
T^{\overline{\nu\mu}}\Big]\over {\rm
Tr}\big[\hat{\varrho}\big]}=(\rho+P)g^{\overline{\nu\mu}}-Pu^{\overline\nu} u^{\overline\mu},
\label{emt}
\end{equation}
where $\rho$  is the energy density. 
The first law of thermodynamics guarantees that there is an additional state
function, the entropy density $\sigma$, related to energy density and
pressure by
\begin{equation}
T\sigma=\rho+P.
\end{equation} 
This relation is  equivalent to the more familiar differential relation $TdS=dU+P dV$. 

It is convenient to express the right hand side of Eq. (\ref{SM1})  as the
trace of the covariant density operator (\ref{Z2}). The left hand side  of (\ref{SM1}) can be written
covariantly in terms of the differential surface element (\ref{dS})
using
$\sqrt{-g}\,dx^{\overline{1}}dx^{\overline{2}}
dx^{\overline{3}}/\sqrt{g_{\overline{00}}}=dS_{\alpha}u^{\alpha}$ as an integral over
the free energy density $\Phi$:
\begin{displaymath}
F/\sqrt{g_{\overline{00}}}=\int dS_{\alpha}u^{\alpha}\Phi.
\end{displaymath}
The manifestly covariant statement of Eq. (\ref{SM1}) is
\begin{equation}
\exp\Big\{\!-\beta \!\int\! dS_{\alpha}\Phi u^{\alpha}\Big\}
={\rm Tr}\Big[\exp\Big\{\!-\!\beta\!\int\! dS_{\alpha} T^{\alpha}_{\;\cdot\;\lambda}
u^{\lambda}\Big\}\Big].\label{SM2}
\end{equation}
Now apply this to  two different equilibrium states with infinitessimally different
$\beta$ and $u^{\alpha}$. The difference gives the differential relation
\begin{equation}
d(\beta\Phi u^{\alpha})={{\rm
Tr}\big[\,\hat{\varrho}\,T^{\alpha}_{\;\cdot\;\lambda}\big]
\over {\rm Tr}[\hat{\varrho}\,]}d(\beta u^{\lambda}).
\end{equation}
The differential on  the left hand side is
\begin{displaymath}
d(\beta\Phi)u^{\alpha}+(\beta\Phi) du^{\alpha}.
\end{displaymath}
Using the thermal average of the energy-momentum tensor (\ref{emt}), the right hand side
is
\begin{eqnarray}
&&\big((\rho+P)u^{\alpha}u_{\lambda}-P\delta^{\alpha}_{\lambda}\big)
\big(d\beta u^{\lambda}+\beta du^{\lambda}\big)\nonumber\\
&&\hskip1cm =\rho d\beta u^{\alpha}-P\beta du^{\alpha},\nonumber
\end{eqnarray} where $u_{\lambda}du^{\lambda}=0$ has been used. 
Equating  the left and right  hand sides and noting that $u^{\alpha}$ and $du^{\alpha}$ are
orthogonal vectors,  gives  the two  relations
\begin{subequations}\begin{eqnarray}
\Phi=&&-P\\
d(\beta \Phi)=&&\rho d\beta.
\end{eqnarray}\end{subequations}
The first relation is the  same as Eq. (\ref{P}).
The second relation implies $d\Phi=(\Phi-\rho)dT/T=-(P+\rho)dT/T
=-\sigma dT$. Therefore  $\sigma=-\partial \Phi/\partial T$, which is the same as
Eq. (\ref{S}) but  expressed in terms of densities.

\subsection{Thermal field theory in imaginary time}

It is also interesting to  quantize in imaginary oblique time by letting $x^{\overline{0}}=-i\tau$
with
$0\le \tau\le\beta/\!\sqrt{g_{\overline{00}}}$.  As shown in Eq. (\ref{D>1}) the
thermal Wightman function is analytic for $\tau$ in this region.
Since the Cartesian coordinates are linear combinations of the oblique coordinates,
$x^{\alpha}=A^{\alpha}_{\overline\mu}\,x^{\overline\mu}$, making
$x^{\overline{0}}$ pure imaginary makes some of the Cartesian coordinates complex. 
In other words, going to imaginary time does not commute with the  coordinate
transformations.  

The Euclidean propagator comes from the replacement $p_{\overline{0}}\to
-i\omega_{n}$ in Eq. (\ref{prop1}) where  $\omega_{n}=2\pi n
T\sqrt{g_{\overline{00}}}$ : 
\begin{equation}
D_{E}(p)={-1\over
g^{\overline{00}}\omega_{n}^{2}+2i\,g^{\overline{0j}}\omega_{n}\,
p_{\overline{j}}-g^{\overline{ij}}p_{\overline{i}}p_{\overline{j}}+m^{2}}.
\end{equation}
Note that the denominator has an imaginary part because of the non-diagonal
metric but the real part of the denominator is positive as always. In perturbation theory
the summation/integration over loop momenta is
\[
{T\sqrt{g_{\overline{00}}}\over\sqrt{-g}}\sum_{n=-\infty}^{\infty}
\int{dp_{\overline{1}}dp_{\overline{2}}dp_{\overline{3}}\over (2\pi)^{3}}.\]

One can canonically quantize in  imaginary time, in which case the field operators 
obey equations of motion. Alternatively one can use a
 a Euclidean functional integral \cite{Das,LeBellac,Kapusta} in which case the fields are periodic 
under
$\tau\to\tau+\beta/\sqrt{g_{\overline{00}}}$. The partition function is
\begin{equation}
Z=\int_{\rm periodic}[D\phi]\exp\Big\{\int\! d^{4}x_{E}\,{\cal L}\Big\},
\end{equation}
where  the integration element in four-dimensional Euclidean space-time is  given  by
\[
\int d^{4}x_{E}=\sqrt{-g}\int_{0}^{\beta/\sqrt{g_{\overline{00}}}}
d\tau\int dx^{\overline{1}}dx^{\overline{2}}dx^{\overline{3}}.
\]
Thermodynamics is computed from the
Helmholtz free energy $F(T,V)$
\begin{equation}
\exp\Big\{\!\!-\beta F/\sqrt{g_{\overline{00}}}\Big\}
=Z
\end{equation}
using Eq. (\ref{thermo}).

\section{Examples}

It is easy to see that time-independent transformations will
not give anything new. More specifically, if $x^{0}=x^{\overline{0}}$ then the
quantization is at fixed $x^{0}$  and if $x^{1}$, $x^{2}$, and
$x^{3}$ are independent of $x^{\overline{0}}$ then $A^{j}_{\overline{0}}=0$ so that
$u^{j}=0$ and the density operator will be $\exp\{-\beta P_{0}\}$. 

This section will deal with a general class of examples based on transformations  of the form
\begin{subequations}\begin{eqnarray}
x^{\overline{0}}=&& ax^{0}+bx^{3}\nonumber\\
x^{\overline{3}}=&& cx^{0}+dx^{3},\label{T1}
\end{eqnarray}
with $x^{\overline{1}}=x^{1}$ and  $x^{\overline{2}}=x^{2}$ always
understood. All the examples in this section will result from special choices of
$a,b,c,d$.
If
$|a|>  |b|$ and $|d|> |c|$
then $x^{\overline{0}}$ is a true time coordinate and
$x^{\overline{3}}$ is a true space coordinate. Light front coordinates
violate this, viz. $|a|=|b|$ and $|c|=|d|$. 
The inverse transformation to (\ref{T1}) is
\begin{eqnarray}
x^{0}=&& (dx^{\overline{0}}-bx^{\overline{3}})/N\nonumber\\
x^{3}=&& (-cx^{\overline{0}}+ax^{\overline{3}})/N,\label{T2}
\end{eqnarray}
where $N=ad-bc\neq 0$.  The two conditions
\begin{equation}
a> 0,\hskip1cm 
{d\over N}>  0,\label{T3}
\end{equation}\end{subequations}
guarantee that increasing values of $x^{\overline{0}}$ corresponds to
increasing
$x^{0}$. 

\paragraph*{(1) Metric.}
The contravariant metric is
\begin{subequations}\begin{equation}
g^{\overline{\mu}\,\overline{\nu}}
=\left(\begin{array}{cccc} a^{2}-b^{2} & 0 & 0 & ac-bd\\
0 & -1 & 0 & 0\\
0 & 0 & -1 & 0\\
ac-bd& 0 & 0 &c^{2}-d^{2}
\end{array}\right),\end{equation}
and the
 covariant metric is
\begin{equation}
g_{\overline{\mu}\,\overline{\nu}}
=\left(\begin{array}{cccc} d^{2}-c^{2} & 0 & 0 & ac-bd\\
0 & -N^{2} & 0 & 0\\
0 & 0 & -N^{2} & 0\\
ac-bd& 0 & 0 & b^{2}-a^{2}
\end{array}\right){1\over N^{2}}.
\end{equation}\end{subequations}
The necessary requirement $g_{\overline{00}}\neq 0$ implies $|d|\neq |c|$;
however, there is nothing wrong with choosing  $|a|=|b|$. 
The $3\times 3$ matrix defined in Eq. (\ref{gamma}) is
\begin{equation}
\gamma_{\overline{ij}}
=\left(\begin{array}{ccc} 
 1 & 0 & 0\\
 0 & 1 & 0\\
0 & 0 & (d^{2}-c^{2})^{-1}
\end{array}\right).
\end{equation}

\paragraph*{(2) Momenta.} The contravariant form of the oblique momenta are $p^{\overline\nu}
=A^{\overline\nu}_{\alpha}\,p^{\alpha}$ in  parallel with Eq. (\ref{T1}). From  this, the covariant
momenta are obtained by applying the metric,
$p_{\overline\mu}=g_{\overline{\mu\nu}}\,p^{\overline\nu}$, with the result
\begin{eqnarray}
p_{\overline{0}}=&& (dp^{0}+cp^{3})/N\nonumber\\
p_{\overline{3}}=&& -(bp^{0}+ap^{3})/N,
\end{eqnarray}
and, of course, $p_{\overline{1}}=-p^{1}$, $p_{\overline{2}}=-p^{2}$.

\paragraph*{(3) Density operator.} The time evolution operator is
\[
P_{\overline{0}}={\partial x^{\alpha}\over
\partial x^{\overline{0}}}P_{\alpha}
={dP_{0}-cP_{3}\over N},\]
and therefore the density operator is
\[
\exp\big\{-\beta
P_{\overline{0}}\,/\!\sqrt{g_{\overline{0}\,\overline{0}}}\,\big\}
= \exp\bigg\{-\beta{|d|P_{0}-cP_{3}\epsilon(d)\over
\sqrt{d^{2}-c^{2}}}\bigg\},\]
after using Eq. (\ref{T3}).  Here $\epsilon(d)=\pm 1$ is the sign function.
As expected,  
 $|d|=|c|$ is  excluded. 
One can understand the form of the density operator in a more physical
way. Since the heat bath is at rest in the $x^{\overline{3}}$
coordinate, its laboratory velocity is $v=dx^{3}/dx^{0}=-c/d$ from 
Eq. (\ref{T1}). Using $\gamma=(1-v^{2})^{-1/2}$ gives the four-velocity
\begin{equation}
U^{\alpha}=\big(\gamma, 0,0,\gamma v\big)
=\bigg({|d|\over\sqrt{d^{2}-c^{2}}},
0,0,{-c\,\epsilon(d)\over\sqrt{d^{2}-c^{2}}}
\bigg)
\end{equation}
Thus the density operator is $\hat{\varrho}=\exp(-\beta P_{\alpha}U^{\alpha})$. 

\subsection{Lorentz transformations and generalizations}

\paragraph*{(1) Quantization at rest but moving heat bath.}
The most intuitive situation physically is to quantize conventionally
at fixed $x^{0}$ but for the heat bath to have a velocity $v$. 
This is easily accomplished by the choice
 $a= 1  , b=0, c=\gamma v, d=\gamma$ so that
\begin{eqnarray}
x^{\overline{0}}=&& x^{0}\nonumber\\
x^{\overline{3}}=&& \gamma(x^{3}+vx^{0}).
\end{eqnarray}
The density operator is 
$\exp\big\{-\beta \gamma(P_{0}-vP_{3})\big\}$.

\paragraph*{(2) Quantization surface and  heat bath  moving
differently.}
A more general option is to  give the quantization surface a velocity $v'$ and
the heat bath a velocity $v$:
\begin{eqnarray}
x^{\overline{0}}=&& \gamma^{\prime}(x^{0}+v'x^{3})\nonumber\\
x^{\overline{3}}=&& \gamma(x^{3}+vx^{0})
\end{eqnarray}
However, only the velocity $v$ enters into the density operator:
$\hat{\varrho}=\exp\big\{-\beta \gamma( P_{0}-vP_{3})\big\}$. Note that the metric will depend
on $v$ and $v'$.

\subsection{Light front and generalizations}

As stated previously, strict light front coordinates in which $|d|=|c|$ are forbidden. 

\paragraph*{(1) Front moving at $v<1$.} An interesting case is the transformation
\begin{eqnarray}
x^{\overline{0}}=&& x^{0}\cos(\theta/2)+x^{3}\sin(\theta/2)\nonumber\\
x^{\overline{3}}=&& x^{0}\sin(\theta/2) -x^{3}\cos(\theta/2),
\end{eqnarray}
where $-\pi/2<\theta<\pi/2$. These  would  be light front coordinates if $\theta$ were allowed to
take  the value $\pm\pi/2$.  
The covariant and contravariant metrics are equal:
\begin{equation}
g_{\overline{\mu}\,\overline{\nu}}
=\left(\begin{array}{cccc} \cos\theta & 0 & 0 & \sin\theta\\
0 & -1 & 0 & 0\\
0 & 0 & -1 & 0\\
\sin\theta & 0 & 0 & -\cos\theta
\end{array}\right)=g^{\overline{\mu\nu}}.
\end{equation}
The density operator is 
\begin{equation}
\hat{\varrho}=\exp\bigg\{-\beta
{P_{0}\cos(\theta/2)+P_{3}\sin(\theta/2)
\over\cos\theta}\bigg\},\end{equation}
and obviously fails at $\theta=\pi/2$. 
Alternatively, one can use $v=\tan(\theta/2)$ and express the
transformation as
\begin{eqnarray}
x^{\overline{0}}= && (x^{0}+vx^{3})/\sqrt{1+v^{2}}\nonumber\\
x^{\overline{3}}= && (vx^{0}-x^{3})/\sqrt{1+v^{2}},
\end{eqnarray}
so that the density operator is
$\exp\big\{-\beta \gamma(P_{0}+vP_{3})\big\}.$

\paragraph*{(2)  Choice of Alves, Das, \& Perez.}
The calculations  in    \cite{ADP}
can be  stated as the choice $a=b=d=1, c=0$:
\begin{eqnarray}
x^{\overline{0}}=&& x^{0}+x^{3}\nonumber\\
x^{\overline{3}}=&& x^{3}\label{mlf}.
\end{eqnarray}
The density operator becomes $\exp\{-\beta P_{0}\}$. The covariant metric is
\begin{equation}
g_{\overline{\mu}\,\overline{\nu}}
=\left(\begin{array}{rrrr} 1 & 0 & 0 & -1\\
0 & -1 & 0 & 0\\
0 & 0 & -1 & 0\\
-1& 0 & 0 & 0
\end{array}\right),
\end{equation}
and so  the covariant coordinates are
$x_{\overline{0}}=g_{\overline{0\nu}}x^{\overline\nu}=x^{0}$
and
$x_{\overline{3}}=g_{\overline{3\nu}}x^{\overline\nu}=-(x^{0}+x^{3})
=-\sqrt{2}x^{+}$. The corresponding covariant momentum components are
\begin{eqnarray}
p_{\overline{0}}=&&p^{0}\nonumber\\
p_{\overline{3}}=&&-(p^{0}+p^{3})= -\sqrt{2}p^{+}.\nonumber
\end{eqnarray}
The contravariant metric is
\begin{equation}
g^{\overline{\mu}\,\overline{\nu}}
=\left(\begin{array}{rrrr} 0 & 0 & 0 & -1\\
0 & -1 & 0 & 0\\
0 & 0 & -1 & 0\\
-1 & 0 & 0 &-1
\end{array}\right).
\end{equation}
Because  $g^{\overline{00}}$ vanishes,
the momentum space propagator is linear in  $p_{\overline{0}}$:
\begin{eqnarray}
g^{\overline{\mu\nu}}p_{\overline\mu}p_{\overline\nu}=&&
-2p_{\overline{0}}p_{\overline{3}}-(p_{\overline{1}})^{2}-
(p_{\overline{2}})^{2}-(p_{\overline{3}})^{2}\nonumber\\
=&& 2\sqrt{2}p^{0}p^{+}-(p^{1})^{2}-(p^{2})^{2}
-2(p^{+})^{2}.\nonumber
\end{eqnarray}
In the Euclidean formulation the contravariant time becomes negative, imaginary:
$x^{\overline{0}}\to -i\tau$; the covariant energy becomes discrete, imaginary:
$p_{\overline{0}}\to -i\omega_{n}$ with $\omega_{n}=2\pi n T$. (Note $g_{\overline{00}}=1$.)
The Euclidean propagator used in \cite{ADP} is
\begin{equation}
{1\over 2i\sqrt{2}\,\omega_{n}\,p^{+}-(p^{1})^{2}-(p^{2})^{2}
-2(p^{+})^{2}-m^{2}}.
\end{equation}

\paragraph*{(3) ADP with moving heat bath.}

It is simple to modify the previous case to allow for a
moving heat bath. Choose $ a=b=1$, $c=\gamma v$, and $d=\gamma$ so that
\begin{eqnarray}
x^{\overline{0}}=&& x^{0}+x^{3}\nonumber\\
x^{\overline{3}}= && \gamma(v x^{0}+x^{3}).
\end{eqnarray}
The density operator is 
$\exp\big\{-\beta \gamma( P_{0}-vP_{3})\big\}$,
corresponding to a moving heat bath. 
As before $g^{\overline{00}}=0$ but now $g_{\overline{00}}=\sqrt{1+v}/\sqrt{1-v}$.

\section{Conclusions}

In standard light front quantization  $g_{\overline{00}}
=g_{++}=0$ and  this makes it impossible  to formulate statistical mechanics and thermal field 
theory. Physically, the problem is the infinite velocity of the light front.

The most interesting possibility is to choose oblique coordinates which satisfy
$g_{\overline{00}}\neq 0$ but 
$g^{\overline{00}}=0$ as in Eq. (\ref{mlf}), which is the case studied by Alves, Das, and Perez
\cite{ADP}. The advantage  of choosing $g^{\overline{00}}=0$ is  that  the denominator of the
propagator,
$g^{\overline{\mu\nu}}p_{\overline\mu}p_{\overline\nu}-m^{2}$, will be linear in the energy
variable $p_{\overline{0}}$. Consequently the propagator will have only one pole and not two. 
This reduces the computational effort required for multiloop diagrams.  Any diagram for which the
kinematics allows 
$N$ propagators to be on shell  would normally produce $2^{N}$ contributions.  However if 
$g^{\overline{00}}=0$,  there will be only one contribution.

A very straightforward application would be to compute the quark and gluon propagators
in the hard thermal loop approximation using Eq. (\ref{mlf}) and verify the rotational invariance of
the dispersion relations \cite{LeBellac}.   A more ambitious task would be to  compute the vertex
functions in the hard thermal loop approximation.

\begin{acknowledgments}

It is a pleasure to thank Ashok Das for sending me his paper \cite{ADP} and 
thereby arousing my interest in the subject.
This work was supported in part by a grant from the U.S. National 
Science Foundation under grant PHY-0099380.

\end{acknowledgments}

\appendix

\section{Lorentz invariance of $\lowercase{g}_{\overline{\mu\nu}}$}

The general non-diagonal metric $g_{\overline{\mu\nu}}$ is always invariant under three
rotations and three Lorentz boosts. However the representation of these six transformations
depend on the coordinate system $x^{\overline\mu}$. 

The usual representation of a Lorentz transformation from one set of Cartesian coordinates to
another, $x^{\alpha'}=\Lambda^{\alpha'}_{\alpha}x^{\alpha}$, leaves invariant the Minkowski
metric tensor:
\begin{equation}
\Lambda^{\alpha'}_{\alpha}\,\Lambda^{\beta'}_{\beta}\,g_{\alpha'\beta'}=g_{\alpha\beta}.
\end{equation}
As  before, the index notation of Schouten \cite{Sc} is  used. Here $\alpha'$
runs over $0', 1', 2', 3'$. The Minkowski metric is invariant:
$g_{0'0'}=g_{00}=1$,
$g_{1'1'}=g_{11}=-1$, etc.

Each Lorentz transformation of the Cartesian coordinates induces a Lorentz transformation of the
oblique coordinates, $x^{\overline\mu}=W^{\overline\mu}_{\overline\rho}x^{\overline\rho}$, where
\begin{equation}
W^{\overline\mu}_{\overline\rho}=A^{\overline\mu}_{\alpha'}\,\Lambda^{\alpha'}_{\alpha}\,
A^{\alpha}_{\overline\rho}
\end{equation}
Because $\Lambda$ keeps the Cartesian metric invariant, $W$ automatically keeps the oblique metric
invariant:
\begin{equation}
W^{\overline\mu}_{\overline\rho}\,W^{\overline\nu}_{\overline\sigma}\, g_{\overline{\mu\nu}}
=g_{\overline{\rho\sigma}}\,.
\end{equation}

\section{Quantization in Oblique coordinates}

This section will perform the explicit quantization in an
arbitrary oblique coordinate system for the free scalar field and then calculate the thermal
average of the free energy-momentum tensor.

\subsection{Equation of motion}

The action expressed as an integral over
contravariant coordinates is
$\sqrt{-g}\int dx^{\overline{0}}dx^{\overline{1}}dx^{\overline{2}}dx^{\overline{3}}\, {\cal L}$
with Lagrangian density
\begin{equation}
{\cal L}={1\over 2}g^{\overline{\mu\nu}}
{\partial\phi\over\partial x^{\overline\mu}}
{\partial\phi\over\partial x^{\overline\nu}}
-{1\over 2}m^{2}\phi^{2}.
\end{equation}
The field equation that follows from the Lagrangian density is 
\begin{equation}
g^{\overline{\mu\nu}}
{\partial^{2}\phi\over\partial x^{\overline\mu}
\partial x^{\overline\nu}}
=m^{2}\phi.
\end{equation}
The solution to this will be a superposition of plane waves of the form 
$\exp\big(-ip_{\overline\alpha}x^{\overline\alpha}\big)$,
with the phase expressed in terms of contravariant spatial coordinates and covariant
momentum coordinates.
The equation of motion  gives the mass shell condition
\begin{equation}
g^{\overline{\mu\nu}}p_{\overline\mu}p_{\overline\nu}=m^{2}.
\end{equation}
This is a quadratic equation for $p_{\overline{0}}$ with two
solutions
\begin{equation}
p_{\overline{0}\pm}=-{g^{\overline{0j}}p_{\overline{j}}
\over g^{\overline{0}\,\overline{0}}}
\pm \bigg[{m^{2}+p_{\overline{i}}p_{\overline{j}}
\Gamma^{\overline{ij}}\over
g^{\overline{00}}} 
\bigg]^{1/2}\label{disprel1}
\end{equation}
where $\Gamma^{\overline{ij}}$ is the $3\times 3$ matrix
\begin{equation}\Gamma^{\overline{ij}}
={g^{\overline{0i}}g^{\overline{0j}}\over g^{\overline{00}}}
-g^{\overline{ij}}.
\end{equation}
The two solutions are
$\exp(-ip_{\overline{0}\pm}x^{\overline{0}}-ip_{\overline{j}}x^{\overline{j}})$. The
energies $p_{\overline{0}\pm}$ are not invariant under momentum inversion, but rather
$p_{\overline{0}-}\to -p_{\overline{0}+}$ when
$p_{\overline{j}}$ changes sign.  Therefore one can use for the second plane wave the
negative momentum solution 
$\exp(ip_{\overline{0}+}x^{\overline{0}}+ip_{\overline{j}}x^{\overline{j}})$.
The solution to the field equation can be expanded as
\begin{equation}
\phi(x)=\!\int\!{dp_{\overline{1}}dp_{\overline{2}}dp_{\overline{3}}\over
\sqrt{-g}(2\pi)^{3}2|p^{\overline{0}}|}\bigg[a(p)e^{-ip\cdot x}
+a(p)^{\dagger}e^{ip\cdot x}\bigg].
\end{equation}
where $p\cdot x=p_{\overline{0}+}x^{\overline{0}}+p_{\overline{j}}x^{\overline{j}}$.
The contravariant energies $p^{\overline{0}\pm}$ are equal in magnitude:
\begin{eqnarray}
p^{\overline{0}\pm}=&&
g^{\overline{00}}p_{\overline{0}\pm}+g^{\overline{0j}}p_{\overline{j}}
\nonumber\\
=&&\pm\sqrt{g^{\overline{00}}}\,\bigg[m^{2}+p_{\overline{i}}p_{\overline{j}}
\Gamma^{\overline{ij}}
\bigg]^{1/2}\label{disprel2},
\end{eqnarray}
and $|p^{\overline{0}}|$ will be denoted simply by $p^{\overline{0}}$ \cite{com1}.

\subsection{Canonical quantization}

For  quantization on the surfaces of constant $x^{\overline{0}}$,
 the canonical momentum is
\begin{equation}
\pi(x)= {\partial {\cal L}\over
\partial\big(\partial \phi/\partial
x^{\overline{0}}\,\big)}=g^{\overline{0\mu}}\,{\partial\phi\over\partial
x^{\overline{\mu}}} ={\partial\phi\over\partial x_{\overline{0}}}
\end{equation}
The explicit mode expansion is
\begin{displaymath}
\pi(x)=-i\int{dp_{\overline{1}}dp_{\overline{2}}dp_{\overline{3}}\over
\sqrt{-g}(2\pi)^{3}2}\bigg[a(p)e^{-ip\cdot x}
-a(p)^{\dagger}e^{ip\cdot x}\bigg]
\end{displaymath}
If the mode operators are required to satisfy
\begin{equation}
\big[a(p),a^{\dagger}(p')\big]
=\sqrt{-g}2\overline{p}^{0}(2\pi)^{3}\prod_{j=1}^{3}\delta(p_{\overline{j}}
-p_{\overline{j}}^{\prime}),
\end{equation}
then equal time commutator has the correct value:
\begin{eqnarray}
&&\big[\pi(x), \phi(x')\big]_{x^{\overline{0}}
=x^{\prime\overline{0}}}\nonumber\\
&&=-i\int{dp_{\overline{1}}dp_{\overline{2}}dp_{\overline{3}}\over
(2\pi)^{3}2}\,
\Big[e^{-ip_{\overline{j}}(x^{\overline{j}}-x^{\prime\overline{j}})}
+e^{ip_{\overline{j}}(x^{\overline{j}}-x^{\prime\overline{j}})}
\Big]\nonumber\\
&&= -{i\over\sqrt{-g}}\,\delta(x^{\overline{1}}-x^{\prime\overline{1}})
\delta(x^{\overline{2}}-x^{\prime\overline{2}})
\delta(x^{\overline{3}}-x^{\prime\overline{3}}).\nonumber
\end{eqnarray}

\subsection{Microcausality}

It is easy to verify microcausality. The commutator of two fields is
\begin{displaymath}
\big[\phi(x),\phi(0)\big]=\int{dp_{\overline{1}}dp_{\overline{2}}dp_{\overline{3}}\over
\sqrt{-g}(2\pi)^{3}2p^{\overline{0}}}
\Big[e^{-ip_{\overline\alpha}x^{\overline\alpha}}
-e^{ip_{\overline\alpha}x^{\overline\alpha}}\Big].
\end{displaymath}
Change to Minkowski integration variables by defining
$p_{\overline\alpha} =p_{\lambda}\partial x^{\lambda}/\partial
x^{\overline{\lambda}}$ so that
$p_{\overline\alpha}x^{\overline\alpha}=p_{\lambda}x^{\lambda}$.  The
integration measure is invariant and therefore
\begin{displaymath}
\big[\phi(x),\phi(0)\big]=\int{dp_{1}
dp_{2}dp_{3}\over(2\pi)^{3}2p^{0}}
\Big[e^{-ip\cdot x}-e^{ip\cdot x}\Big].
\end{displaymath}
This is the conventional answer for the commutator. It vanishes for space-like separations
$x_{\lambda}x^{\lambda}<0$. Since
$x_{\overline\alpha}x^{\overline\alpha}=x_{\lambda}x^{\lambda}$ it
vanishes for $x_{\overline\alpha}x^{\overline\alpha}<0$.

\subsection{Hamiltonian}

The canonical Hamiltonian density is
\begin{equation}  
{\cal H}=\pi{\partial\phi\over\partial x^{\overline{0}}}-
{\cal L}\label{Hdensity}
\end{equation}
It is convenient to express this in terms of mixed contravariant and covariant derivatives:
\begin{displaymath}
{\cal H}\!=\!{1\over 2}\bigg[
{\partial\phi\over\partial x_{\overline{0}}}
{\partial\phi\over\partial x^{\overline{0}}}
\!-\!{\partial\phi\over\partial x_{\overline{1}}}
{\partial\phi\over\partial x^{\overline{1}}}
\!-\!{\partial\phi\over\partial x_{\overline{2}}}
{\partial\phi\over\partial x^{\overline{2}}}
\!-\!{\partial\phi\over\partial x_{\overline{3}}}
{\partial\phi\over\partial x^{\overline{3}}}
\!+\!m^{2}\phi^{2}\bigg]
\end{displaymath}
The Hamiltonian requires integrating over the contravariant three-volume
\begin{displaymath}
P_{\overline{0}}=\sqrt{-g}\int
dx^{\overline{1}}dx^{\overline{2}}dx^{\overline{3}}\;
{\cal H}.
\end{displaymath}
Working this out explicitly gives
\begin{equation}
P_{\overline{0}}=\int{dp_{\overline{1}}dp_{\overline{2}}dp_{\overline{3}}\over
\sqrt{-g}(2\pi)^{3}2p^{\overline{0}}}\;{p_{\overline{0}}\over 2}
\Big[a^{\dagger}(p)a(p)+
a(p)a^{\dagger}(p)\Big].
\end{equation}
Note that the covariant energy $p_{\overline{0}}$ in the numerator does
not cancel the contravariant energy  $p^{\overline{0}}$ in the
denominator. The commutation relation
\begin{displaymath}
\big[\,P_{\overline{0}},\phi(x)\big]=
-i{\partial\phi\over\partial x^{\overline{0}}}
\end{displaymath}
verifies that the Hamiltonian is the generator of translations in the contravariant
time variable $x^{\overline{0}}$.

\subsection{Energy and  momentum}

The canonical energy-momentum tensor is
\begin{equation}
T^{\overline\nu}_{\,\cdot\,\overline\mu}={\partial\phi\over\partial
x_{\overline\nu}}{\partial\phi\over\partial
x^{\overline\mu}}
-\delta_{\overline\mu}^{\overline\nu}\,{\cal L}
\label{em1}
\end{equation}
and satisfies the conservation laws $\partial
T_{\;\cdot\;\overline\mu}^{\overline\nu}/\partial
x^{\overline\nu}=0$.  The $\overline\mu=\overline{0}$ and $\overline\mu=\overline{m}$
components of this equation are
\begin{eqnarray}
0=&& {\partial T_{\;\cdot\;\overline{0}}^{\overline{0}}\over\partial
x^{\overline{0}}} +{\partial T_{\;\cdot\;\overline{0}}^{\overline{n}}\over\partial
 x^{\overline{n}}}\\
0=&& {\partial T_{\;\cdot\;\overline{m}}^{\overline{0}}\over\partial x^{\overline{0}}}
+{\partial T_{\;\cdot\;\overline{m}}^{\overline{n}}\over\partial x^{\overline{n}}}
\end{eqnarray}
From  Eq. (\ref{Hdensity}), 
${\cal H}=T_{\;\cdot\;\overline{0}}^{\overline{0}}$ and thus
$T_{\;\cdot\;\overline{0}}^{\overline{0}}$ is the energy density.  The first of the
above equations indentifies $T_{\;\cdot\;\overline{0}}^{\overline{n}}$ as the energy
flux. From the second, 
$T_{\;\cdot\;\overline{m}}^{\overline{0}}$ is the momentum density and
$T_{\;\cdot\;\overline{m}}^{\overline{n}}$ is the momentum flux.
Integrating the energy and momentum densities over a contravariant three-volume gives
\begin{equation}
P_{\overline\mu}=\sqrt{-g}\int
dx^{\overline{1}}dx^{\overline{2}}dx^{\overline{3}}\;
T^{\overline{0}}_{\,\cdot\,\overline{\mu}}.
\end{equation}  
These integrals are independent of the contravariant time:
$\partial P_{\overline\mu}/\partial x^{\overline{0}}=0$.
They generate translations in the contravariant coordinates:
\begin{equation}
\big[\,P_{\overline\mu},\phi(x)\big]=-i{\partial\phi\over\partial x^{\overline\mu}}.
\end{equation}
The explicit form for the three-momentum operators is
\begin{equation}
P_{\overline{j}}=\int {dp_{\overline{1}}dp_{\overline{2}}dp_{\overline{3}}\over
\sqrt{-g}(2\pi)^{3}2p^{\overline{0}}}\;{p_{\overline{j}}
\over 2}
\Big[a^{\dagger}(p)a(p)
+a(p)a^{\dagger}(p)\Big].
\end{equation}

\subsection{Thermal Averages}

This section will show that despite the somewhat complicated dispersion relations
in Eqs. (\ref{disprel1}) and (\ref{disprel2}), the thermal average of the
energy-momentum tensor can be computed directly and gives the conventional answer.

Bose-Einstein statistics gives for the thermal
average of the energy-momentum tensor of a free gas of scalar particles
\begin{equation}
\langle
T_{\overline{\mu\nu}}\rangle=\int\!{dp_{\overline{1}}dp_{\overline{2}}
dp_{\overline{3}}\over \sqrt{-g}(2\pi)^{3} p^{\overline{0}}}
\;{p_{\overline\mu}p_{\overline{\nu}}\over
\exp\big(\beta p_{\overline{0}}/\!\sqrt{g_{\overline{00}}}\big)-1}.
\end{equation}  
To perform this integration, change to Minkowski momenta $k_{\alpha}$, where
\begin{equation}
p_{\overline\mu}=A^{\alpha}_{\overline\mu}\,k_{\alpha}.
\end{equation}
The mass shell condition requires $k_{0}=(\bm{k}^{2}+m^{2})^{1/2}$.
And $p_{\overline{0}}/\!\sqrt{g_{\overline{00}}}=k_{\lambda}u^{\lambda}$ with
$u^{\lambda}=A^{\lambda}_{\overline{0}}\,u^{\overline{0}}$, the oblique velocity 
given by Eq. (\ref{vel}). The change of variables gives
\begin{equation}
\langle T_{\overline{\mu\nu}}\rangle
=A^{\alpha}_{\overline\mu}A^{\beta}_{\overline\nu}\int{dk_{1}dk_{2}dk_{3}
\over (2\pi)^{3}k_{0}}{k_{\alpha}k_{\beta}\over
\exp(\beta k_{\lambda}u^{\lambda})-1},
\end{equation}
whose evaluation is standard:
\begin{eqnarray}
\langle T_{\overline{\mu\nu}}\rangle
=&&A^{\alpha}_{\overline\mu}A^{\beta}_{\overline\nu}
\Big((\rho+P)u_{\alpha}u_{\beta}-Pg_{\alpha\beta}\Big)\nonumber\\
=&& (\rho+P)u_{\overline\mu}u_{\overline\nu}-Pg_{\overline{\mu\nu}}.
\end{eqnarray}
The final result is expressed in terms of the oblique velocity vector and the oblique metric
tensor. The most physical quantity is the mixed tensor
\begin{equation}
\langle T^{\overline\mu}_{\;\cdot\;\overline{\nu}}\rangle
=\left(\begin{array}{cccc} \rho & 0 & 0 & 0
\\ (\rho\!+\!P)g_{\overline{01}}/g_{\overline{00}}  & -P & 0 & 0\\
(\rho\!+\!P)g_{\overline{02}}/ g_{\overline{00}}  & 0 & -P & 0\\
(\rho\!+\!P)g_{\overline{03}}/ g_{\overline{00}} & 0 & 0 &-P
\end{array}\right),
\end{equation}
where Eq. (\ref{vel}) has been used. 
 The off-diagonal entries in the metric give a nonzero momentum density:
$\langle
T^{\overline{0}}_{\;\cdot\;\overline{n}}\rangle=(\rho+P)g_{\overline{0n}}/g_{\overline{00}}$.


\begin{thebibliography}{}
\bibitem{com1} If  $g_{\overline{00}}$ were negative,
 the energies $p_{\overline{0}}$ and  $p^{\overline{0}}$ would be 
complex for certain real momentum.
\end{thebibliography}
\end{document}